# Policy Aware Geospatial Data

Puneet Kishor
Nelson Institute for Environmental Studies
University of Wisconsin-Madison
1630 Linden Drive,
Madison, WI 53706
+1 (608) 729-5853
kishor@wisc.edu

Oshani Seneviratne
Computer Science and Artificial Intelligence Laboratory
Massachusetts Institute of Technology
32 Vassar Street,
Cambridge, MA
+1 (617) 324-8410
oshani@csail.mit.edu

Noah Giansiracusa
Department of Mathematics

Brown University
151 Thayer St., Box 1917
Providence, RI 02912
+1 (401) 863-9928
noahgian@math.brown.edu

## ABSTRACT
Digital Rights Management (DRM) prevents end-users from using content in a manner inconsistent with its creator's wishes. The license describing these use-conditions typically accompanies the content as its metadata. A resulting problem is that the license and the content can get separated and lose track of each other. The best metadata have two distinct qualities – they are created automatically without user intervention, and they are embedded within the data that they describe. If licenses are also created and transported this way, data will always have licenses, and the licenses will be readily examinable. When two or more datasets are combined, a new dataset, and with it a new license, are created. This new license is a function of the licenses of the component datasets and any additional conditions that the person combining the datasets might want to impose. Following the notion of a data-purpose algebra, we model this phenomenon by interpreting the transfer and conjunction of data as inducing an algebraic operation on the corresponding licenses. When a dataset passes from one source to the next its license is transformed in a deterministic way, and similarly when datasets are combined the associated licenses are combined in a non-trivial algebraic manner. Modern, computer-savvy, licensing regimes such as Creative Commons allow writing the license in a special kind of language called Creative Commons Rights Expression Language (ccREL). ccREL allows creating and embedding the license using RDFa utilizing XHTML. This is preferred over DRM which includes the rights in a binary file completely opaque to nearly all users. The colocation of metadata with human-visible XHTML makes the license more transparent. In this paper we describe a methodology for creating and embedding licenses in geographic data utilizing ccREL, and programmatically examining embedded licenses in component datasets and determining the resulting license of the composite dataset as determined by the relevant data-purpose algebra. We are inspired by the concept of affordance as it applies in the context of human-computer interaction (HCI). Instead of using technology to make it difficult for the user to do the wrong thing, we want to use technology to make it easy for the user to do the right thing. A technical solution that will assist the user do the right thing can go a long way in easing the burden on the authors creating and distributing licenses along with data, and in easing the burden on the users determining the appropriate use of datasets based on their licenses. This can assist in implementing a policy that protects intellectual property while encouraging sharing and use.

## Categories and Subject Descriptors
E.0 [**Data General**]: Legal Aspects

## 1. INTRODUCTION
Data providers typically attach a license to their data describing what users may or may not do with that data. Such a license is typically in the form of legal text either printed on a sheet of paper, or available as HTML from a link on the web. In that sense, license is really metadata that describes something about the data.

### 1.1. The Problem
We identify four problems with the typical methods of conveying a license:

**Laziness:** We are inherently lazy, wanting to do the least possible to achieve the most. Already busy with working with our data, we usually can't be bothered to tinker with metadata.

**Separation:** Even if metadata are created, they can and do get easily separated from the data. This is akin to having a name tag that, instead of being pinned to our lapels, is kept in a different room; not very useful, and not around when needed.

**Interoperability:** When two or more datasets are mashed up together, the user has to figure out what use permissions the newly created dataset would inherit. While for most metadata, the solution is simple accretion, for licenses this can get tricky because different licenses can be at odds with each other making the datasets incompatible. It must be noted that there are other levels of interoperability besides legal – we want the licenses to also be interoperable at semantic and technological levels.

**Generativeness:** Building upon the interoperability problem, we use the term generativeness to describe the problem pertaining to generation, the ability to produce a new license easily.

These four issues affect licensing, propagation of licenses, and legal and compatible uptake of data at all levels – users don't produce requisite licenses correctly; if they do produce the licenses, the licenses are not embedded inside the data, hence the two get separated, making it difficult for users to determine the data and the related license easily; if they are able to find the license, they are unable to easily determine the license of the new dataset created upon mashing up two or more datasets; and finally, they are unable to create the new license, and the cycle continues.

## 1.2. Discussion

Since a license is really just metadata, this problem becomes, among other things, an extension of the problem of managing the metadata. It is also a behavioral problem, and a legal problem.

There have been many initiatives to embed metadata into or otherwise colocate with the data they describe. Extensible Metadata Platform (XMP) is a technology embeds metadata in machine readable Resource Description Framework (RDF). [1] This technology is widely deployed in embedding licenses in free-floating multimedia content such as images, audio and video on the web. The Exchangeable Image File Format [2] and the IPTC4XMP describe metadata that can be embedded within photographs. Several file systems also allow the use of extended file attributes for associating metadata with the files they describe.

Jones (2007) developed a bookmarklet to copy/paste document fragments while preserving provenance information that was stored inline using XHTML syntax. Building up on that work, Seneviratne has created a Semantic Clipboard based on CC licenses with possible extensions to scenarios modeled in policy languages such as Accountability In RDF (AIR) (Kagal, Hanson, and Weitzner 2008). The Semantic Clipboard is Firefox extension to seamlessly integrate metadata with content upon reuse, capturing intent of the usage and

|  | PD | CC0 | BY | BY NC | BY NC ND | BY NC ND SA | BY NC SA | BY ND | BY ND SA | BY SA | ARR | NL |
|---|---|---|---|---|---|---|---|---|---|---|---|---|
| PD | PD | CC0 | BY | BY NC | BY NC ND | BY NC ND SA | BY NC SA | BY ND | BY ND SA | BY SA | ARR | NL |
| CC0 | CC0 | CC0 | BY | BY NC | BY NC ND | BY NC ND SA | BY NC SA | BY ND | BY ND SA | BY SA | ARR | NL |
| BY | BY | BY | BY | BY NC | BY NC ND | BY NC ND SA | BY NC SA | BY ND | BY ND SA | BY SA | X | X |
| BY NC | BY NC | BY NC | BY NC | BY NC | BY NC ND | BY NC ND SA | BY NC SA | X | X | X | X | X |
| BY NC ND | BY NC ND | BY NC ND | BY NC ND | BY NC ND | BY NC ND | BY NC ND SA | X | BY NC ND | X | X | X | X |
| BY NC ND SA | BY NC ND SA | BY NC ND SA | BY NC ND SA | BY NC ND SA | BY NC ND SA | BY NC ND SA | BY NC ND SA | BY NC ND SA | BY NC ND SA | BY NC ND SA | X | X |
| BY NC SA | BY NC SA | BY NC SA | BY NC SA | BY NC SA | X | BY NC ND SA | BY NC SA | X | X | X | X | X |
| BY ND | BY ND | BY ND | BY ND | X | BY NC ND | BY NC ND SA | X | BY ND | BY ND SA | X | X | X |
| BY ND SA | BY ND SA | BY ND SA | BY ND SA | X | X | BY NC ND SA | X | BY ND SA | BY ND SA | BY ND SA | X | X |
| BY SA | BY SA | BY SA | BY SA | X | X | BY NC ND SA | X | X | BY ND SA | BY SA | X | X |
| ARR | ARR | ARR | X | X | X | X | X | X | X | X | X | X |
| NL | NL | NL | X | X | X | X | X | X | X | X | X | X |

PD: Public Domain  
CC0: Creative Commons Zero  
BY: Attribution  
NC: Non Commercial  
ND: No Derivatives  
SA: Share Alike  
ARR: All Rights Reserved  
NL: No License  
X: The two licenses can't be combined

**Figure 1: License Matrix**

ensuring that the content will be reused in a policy aware manner (Seneviratne 2009). These projects have contributed a clearly defined document fragment ontology that can represent the information about the sources of the content, a method of excerpting from these sources, and a reasoning engine which reasons over the acceptable use of the source and the composite CC licenses.

## 1.3. The Solution

Our proposed solution embeds metadata within the data itself and builds upon the ideas of inline provenance (Jones 2007), rights expression language such as ccREL (Commons 2009) and data-purpose algebra (Hanson et al. 2007).

First, if we generate a license describing the use-conditions automatically, it solves the *laziness* problem. Second, if we embed the license within the data, much like EXIF/IPTC metadata in photographs, it solves the *separation* problem. Third, if we utilize a standard, rights expression language such as ccREL, we solve the *interoperability* problem. And, finally, we programmatically compute the new license for a data set created by mashing-up two or more data sets, thereby solving the *generative* problem.

To generate the license metadata, we are inspired by the "Data-Purpose Algebra." If we can extract the license metadata that have been embedded inline using a rights expression language (REL), then we can combine and calculate new license or *noops* from mashing-up two or more data streams.

---

[1] Adobe, Inc., created the XMP specification.

[2] EXIF and relation resources at http://www.exif.org

## 1.4. License Algebra

Under our proposed framework of viewing usage-licenses as a type of metadata embedded in the datasets that they describe, an issue arises when a dataset is modified or multiple datasets are combined in some manner (we use the term "mash-up" to refer to the process of dynamically combining data) – namely, a new license must be created based on the constituent datasets, the process in which they were combined, and the original licenses embedded within them, and this new license should be embedded in the new dataset following the same formatting and protocol as the metadata associated with the constituent data. In broad terms, we mathematically describe the transformation of metadata resulting from a transformation of the corresponding data.

To illustrate this concept, suppose the U.S. Census Bureau wants to collect demographic information. To distribute the labor of going door-to-door asking people questions, they hire a handful of pollers to collect the information, each one assigned to a different neighborhood. Along with the demographic data, the pollers take note of the number of households that did not answer the door. We think of this latter number as metadata associated with the demographic data itself. When the organization is ready to compile the datasets to produce one large database with all the demographic information, they also need to compile the metadata. In this naive illustration the total number $q$ of households that did not answer the door is simply the sum of the number $q_i$ found by each individual poller, namely $q = q_1 + ... + q_n$ (assuming there are $n$ pollers). In this simplistic example, combining the specified metadata may seem so obvious that it's not worth describing it with such abstract formalism, but generally the process of combining or mashing up metadata can be described algebraically, especially when the metadata is used to describe the intended purpose and/or limit the usage of the associated data – whence the term *data-purpose algebra* (Hanson et al. 2007).

Consider a roads dataset on one website and a restaurants dataset on another website. Each dataset has a license describing its intended use-conditions as designated by the Creative Commons licenses such as No Derivatives (ND), Non Commercial (NC), etc. A developer wishes to create a new web application by mashing up the roads and the restaurants datasets with her own neighborhoods dataset to allow users to find directions to restaurants. What usage designations would be associated with the new application created from merging the three component data? A heuristic approach is to encode each possible license designation (ND, NC, etc.) as a single bit in a bit-sequence of length equal to the total possible number of constituent designations, and then the data-purpose algebra is described by performing a bit-wise logical *OR* on the two input sequences. For instance, if one dataset is labeled as ND and another as NC, then we can encode the licenses as 01 and 10 respectively, so the license on the mash-up with both datasets will be 01+10 = 11, which stands for "ND NC," meaning the data in the new application can be used for non-commercial purposes and no derivative works may be made of the combined data. Hence, the naive "algebraically summing" process is actually quite versatile.

One complication with the above heuristic for combining licenses is that certain combinations of licenses are incompatible (something cannot be "All Rights Reserved" and "By Attribution" at the same time), and some combinations of licenses are redundant (something that is both "All Rights Reserved" and "Public Domain" is *de facto* "All Rights Reserved").

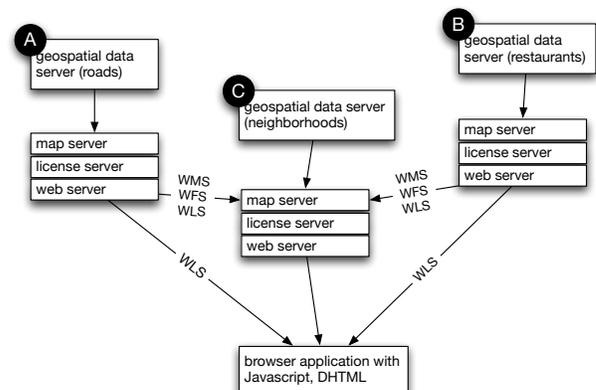

**Figure 2: Web License Service**

The matrix in Figure 1 is an explicit recipe for combining the possible license designations. Mathematically, it describes the structure of the data-purpose algebra for combining licenses, and with it one can readily implement an algorithm for computing licenses. One important observation is while all data-purpose algebras may not be symmetric, our license matrix is symmetric, that is, combining dataset A with B is the same as combining them in the opposite order: $A+B=B+A$. Another property not observable from the matrix but true of nearly all algebras (such as those in this paper) is *associativity*: if "+" denotes the composition law in the algebra, and $A$, $B$, $C$ are various license combinations, then $A+(B+C)=(A+B)+C$. This means that when combining three or more datasets, a new license can be computed by combining the datasets two-at-a-time in any order one chooses.

One of the common formats for geospatial data is the Shapefile format (ESRI 1998). The Shapefile format stores a single feature type in a collection of files, each one of them holding some aspect of the feature information. The main files for a line feature dataset of roads are: roads.shp, the geometry of the lines; roads.dbf, a table of attributes, one row per feature; and roads.shx, a cross-reference between the shp and the dbf files. The Shapefile format prescribes a few optional files such as: roads.prj, the metadata for the projection information; roads.qix, the quadtree index for the dataset; and various spatial and attribute indexes and other files

We propose an additional file called roads.lic that would hold the license metadata for the dataset. This license would be stored using ccREL which allows expressing a licensing utilizing RDFa (W3C 2008a) in XHTML (W3C 2008b). The user can simply go to the CC license chooser, [3] choose a license through its step-by-step set of screens, and copy and save the license information expressed in XHTML as plain text in a roads.lic file.

The ccREL-based license is human-readable, and also points to a location on the web from where more information on the license, and its legal code can be retrieved. This makes ccREL suitable for expressing the license, and since the license is just a plain text file, it can also be stored as a record in a table if the geospatial data are stored in a relational database. This makes the license suitable for two of the commonest geospatial data formats.

---

[3] Choose a License http://creativecommons.org/choose/

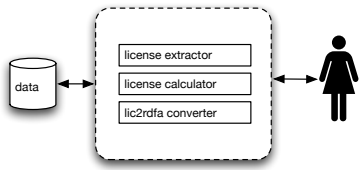

**Figure 3: License Server**

## 1.5. License Server

In a typical application, the user serves the dataset via a map server that works in conjunction with a web server to respond to user requests for data. The map server typically gets its guidance from a configuration file that describes all the data layers in the application, their source, selection parameters, and even their styles including symbology, font, colors and other cartographic elements. The map server queries the data sources, extracts the query results, and typically constructs a map image in a standard image format that is then served by the web server.

As shown in Figure 2, the Open Geospatial Consortium (OGC) describes Web Mapping Service (WMS) and Web Feature Service (WFS) that allow querying map data over HTTP, thus enabling map applications that draw upon distributed data sets along with local data (Consortium 2004) and (Consortium 2005).

We demonstrate a license server analogous to the map server. The license server utilizes a configuration file to query a data set, but instead of querying the geographic features and returning a map image or features, it queries the license information of the data set, processes it, and returns it as an XHTML stream thereby creating a mechanism we call Web License Service (WLS) as shown in Figure 2. The license server (Figure 3) itself is made up of: a license extractor; a license computation engine; and a lic2rdfa converter.

The user requests the license over HTTP, the license extractor extracts the license from the component datasets, the license calculator computes the new license using an efficient license lookup table, the lic2rdfa converter creates an XHTML fragment that expresses the license using ccREL, and the server sends that information back to the user's browser in a machine readable format such as RDFa where that information can be displayed via a link or in a popup window. An example license lookup table is shown in Figure 1. The rows and columns of the table constitute the component licenses, and the corresponding cell gives the resultant license if the two licenses can be combined. When we have more than one data source, we can use the algorithm listed in Figure 4 to recursively query the lookup table and construct the resultant license thereby scaling up to multiple data sources.

```
RECURSIVE-LICENSE-LOOKUP(license, license_list)
1 n = len(license_list)
2 head = license_list[0]
3 tail_list = license_list[1:n]
4 if n == 1    //This is the base case
5     return lookuptable(license, head)
6 else
7     return lookuptable(license,
        RECURSIVE-LICENSE-LOOKUP(head, tail_list))
```

**Figure 4: License Lookup Pseudocode**

## 2. POLICY IMPLICATIONS

Policies in general are pervasive in Web applications. They play a crucial role in enhancing security, privacy and usability of the services offered on the Web (Bonatti et al. 2006). In the realm of data dissemination, access, and sharing, users are encumbered by confusion and opacity in terms of licenses and legal use of data.

The idea of a commons of information has been received with much enthusiasm. The downstream benefits of public sector information can be tremendous to society, especially the possibly unforeseen uses because of widespread access afforded by digital information and computer networks (Uhlir and Schröder 2007).

Not surprisingly, commons has been proposed as a model for geospatial data as well (Onsrud et al. 2004), but it depends on a central repository, and creates many new and unique legal considerations (Mccurry et al. 2006). One proposed solution is to have all the players contractually agree to freely open up their data (Reichman and Uhlir 2003).

We believe that utilizing technology to enable users determine for themselves right versus wrong empowers them instead of alienating them, and is more likely to promote openness and sharing leading to more creativity and economic value added. Technology is more likely to be seen as a non-partisan arbiter of rights, guiding the users to just do the right thing. To summarize, instead of using technology to prevent users from doing the wrong thing, we use technology to help them do the right thing.

### 2.1. Further Work Required

In this paper we have shown a concept. While we have a working demo of the concept, much work remains to be done. First, a more comprehensive demo with comprehensive tests involving various licenses and dataset, storage formats and transportation mechanisms needs to be done. Second, the different alternatives for rights expression have to be evaluated. Finally, a complete and robust specification for the Web License Service has to be developed so various technical client and server solutions can be created by independent programmers and vendors.